\documentclass[aps,twocolumn,showpacs,preprintnumbers,amsmath,amssymb,floatfix]{revtex4-1}

\usepackage{graphicx,tabularx}
\usepackage{dcolumn}
\usepackage{bm}
\usepackage{color}
\usepackage{epstopdf}
\usepackage{upgreek}
\usepackage{color}

\begin{document}


\title{Ordered Defects: A Roadmap towards
room temperature Superconductivity and Magnetic Order}

\author{Pablo D. Esquinazi}
\email{esquin@physik.uni-leipzig.de} \affiliation{Division of
Superconductivity and Magnetism, Felix Bloch Institute for Solid
State Physics, Faculty of Physics and Earth Sciences, University
of Leipzig, Linn\'estrasse 5, D-04103 Leipzig, Germany}

\date{\today}
\begin{abstract}
Defects in the atomic lattice of solids are sometimes
desired. For example,
atomic vacancies, single ones or
 more elaborated defective structures,  can generate localized magnetic moments in a non magnetic
crystalline lattice. Increasing their density to a few percent
magnetic order can appear.  Furthermore, certain two dimensional
interfaces can give rise to localized superconductivity with  a
broad range of critical temperatures. Old and new experimental
facts emphasize the need to  join efforts   to start using
systematically ``ordered defects" in solids to achieve room
temperature superconductivity and magnetic order.
\end{abstract}

\maketitle

Magnetic and superconducting orders at room temperature are highly
desirable due to the large number of possibilities to apply these phenomena in devices
at normal life conditions, apart from the huge basic research
interest. Although magnetic order is found at 300~K  in a not so
large list of materials, superconductivity at room temperature
appeared to be much more difficult to find \cite{som19}. Here, we would like to
emphasize that ordered defects in some lattice structures can
provide us a path to reach both phenomena at very high
temperatures in materials that do not show them in their defect
free state. We would like to pay attention here on two cases of
lattice defects in solids. Namely, a single or  a group of
vacancies and two dimensional (2D) well defined interfaces in
some specific atomic lattices.

Vacancies can trigger a magnetic moment around its position in the
atomic lattice. A large number of experimental and theoretical
work has been done in this respect. For example, STM local
measurements revealed that  C-vacancies, produced by low energy
ion irradiation at the surface of graphite, have a local magnetic
moment \cite{uge10}. Having a large enough density ($\sim 5\%$)
of hydrogen (or protons) or C-vacancies at certain positions
\cite{yaz10,yaz16}, one can show experimentally that magnetic
order at room temperature appears in graphite bulk samples.
Several studies with techniques like element specific X-ray
magnetic circular dichroism (XMCD) \cite{ohldagl,ohldagnjp},  NMR
\cite{fre15},   magnetization and transport (see \cite{chap3} and
refs. therein)  indicate that the magnetic order triggered by
defects is intrinsic and with Curie temperatures clearly above
300~K. In the case of graphite \cite{ohldagnjp} or ZnO
\cite{her10,lor15apl}, XMCD results indicate that the valence band
is spin polarized in a relative  large energy range, an apparently
general feature in materials that show defect induced magnetism
(DIM).

Due to the rather simple way to trigger in non-magnetic materials,
room temperature  magnetic order by low energy ion irradiation, we
may ask whether  some kind of  devices have been already proposed.
Two recent examples are worth  mentioning. The first is the spin
filter that occurs at the interface between magnetic and non
magnetic regions  of the same material, ZnO:Li in the reported
case \cite{bot17}. Whereas the magnetic path of the oxide micro-
or nanostructure is produced by an inexpensive  $\sim 300$~eV
proton irradiation plasma chamber, the protected non-magnetic
semiconducting regions act as a potential well for the thermally
activated conduction electrons. The interfaces between magnetic
and non-magnetic regions do produce a giant positive
magnetoresistance, in contrast to the small and negative
magnetoresistance of the  magnetic paths alone. This
characteristic and other details of the  homo-junctions open up a
new and simple way to use the spin splitting created in the
irradiated oxide for spintronic devices.

Other unexpected result was obtained recently by  low-energy ion
irradiation on TiO$_2$ films. After a gentle ion irradiation
fluence, the originally non-magnetic film becomes magnetic at room
temperature due to Ti-divacancies (which are stable at room
temperature) and with the magnetization vector normal to the main
area of the film \cite{sti16}. The rather large magnetic
anisotropy is related apparently to the fact that the magnetic
layer resides at the very near surface region. Further increase of
the  amount of defects  by subsequent ion irradiation, vanishes
the magnetic anisotropy. Recently obtained results \cite{sti}
indicate that should be possible to produce nanostructured areas
of TiO$_2$ with perpendicular magnetic anisotropy. Low energy ion
irradiation and the existence of DIM in several oxides may open up
a new method to reach perpendicular magnetic anisotropy by far
more simple and economically advantageous than several others used
nowadays \cite{tud17}.
\begin{figure}
\includegraphics[width=\columnwidth]{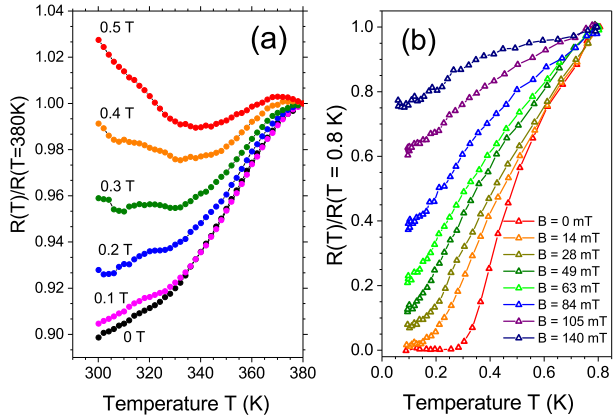}
\caption{\label{tra} Normalized resistance vs. temperature at constant magnetic fields applied normal to
the graphene planes and interfaces of the following samples: (a) Nature graphite sample; data taken from Fig.6(b) in \cite{pre16}.  (b) Bilayer graphene device M2 (twist angle $\theta = 1.05^{\circ}$), data taken from Fig.1(b) in \cite{cao18}.}
\end{figure}

Let us now discuss the other order phenomenon
 that appears at a specially  ordered lattice defect,  namely at certain 2D interfaces.
Systematically done  STM studies of graphene bilayers showed  the
existence of Van Hove singularities in the electronic density of
states that shift to lower bias voltages the smaller the twist
angle between the graphene layers \cite{bri12}. As emphasized
recently by Volovik (see \cite{vol18} and refs. therein) the Van
Hove singularities are related to the flattening of the electronic
energy band at well defined regions   related to the measured
moir\'e pattern in the electronic spectrum.  This appears to be
the reason for the existence of superconductivity found recently
in bilayer graphene with critical temperatures around 1~K
\cite{cao18,mar18}.

We may ask now, whether well  ordered bulk  graphite samples  have  similar 2D interfaces.
The answer is yes, indeed, and the experimental evidence is overwhelming.  Moir\'eŽ patterns in the
electronic spectrum measured by STM due to misoriented graphene layers
 of the graphite structure were found
already in 1990,  at the surface of a highly oriented  pyrolytic
graphite (HOPG) bulk sample \cite{kuw90}, see also \cite{mil10}.
Their influence on the measured conductivity was not, however,
realized till 2008 (for a review see \cite{chap7} and also
\cite{zor18}). There are two more 2D interfaces that can appear
parallel to the graphene planes in real graphite samples, namely:
(a) The one between  regions with Bernal and rhombohedral (RH)
stacking orders, twisted or untwisted, and (b) between twisted RH
regions. The existence of the minority RH stacking phase was
confirmed in a large number of graphite bulk samples  by XRD
studies \cite{pre16} and its influence in the conductivity was
recently reported \cite{ZQSSEKMLL17}.
\begin{figure}
\includegraphics[width=\columnwidth]{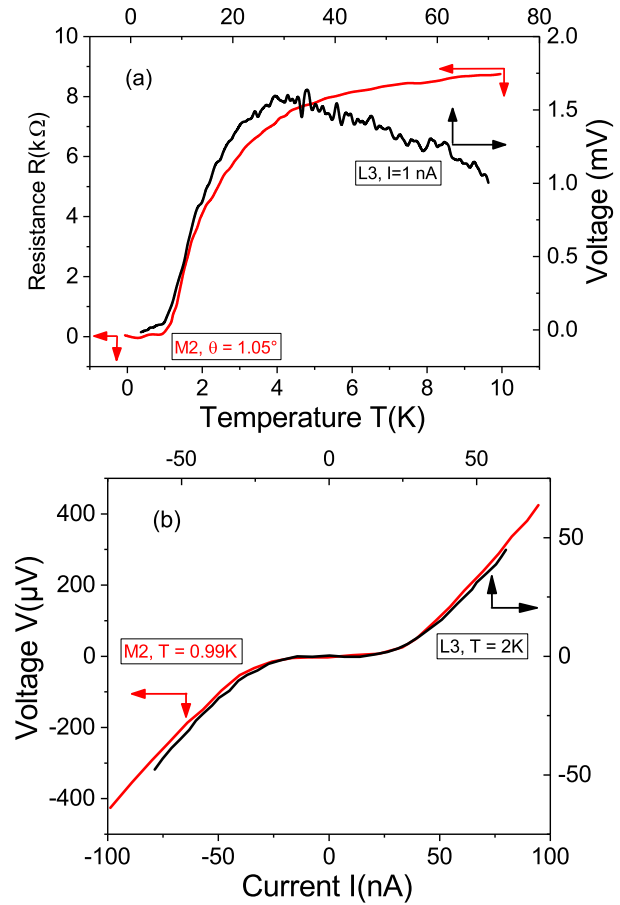}
\caption{\label{lam} (a) Temperature dependence of the resistance or voltage (at constant current)
measured in the bilayer graphene device M2 (red line, taken from
Fig.1(b) in \cite{cao18}, bottom-left axes) and lamella L3 (black line, taken from Fig.3(c) in \cite{bal13}, upper-right axes).
(b) Characteristic voltage-current curves at constant
temperatures obtained in the same samples shown in (a). The data of lamella L3 (black line) were taken from Fig.3(f) in \cite{bal13} (upper-right axes) and from the
bilayer graphene device (red line) from Fig.1(e) in \cite{cao18} (bottom-left axes).}
\end{figure}

Superconductivity at these interfaces is expected  to appear at
very high temperatures due to the existence of a dispersionless
electron band, a so-called flat band \cite{vol18}. This flat band
has been predicted to exist  at, e.g., the surface of graphite
with RH stacking order \cite{kop11,vol18} or at the 2D interface
between Bernal and RH stacking orders \cite{mun13}. Its existence
was confirmed experimentally  at the surface of small and thin RH
patches surrounded by regions with Bernal stacking order
\cite{pie15}.

It is interesting to compare  the superconducting transitions
identified in the temperature dependence  of the electrical
resistance at constant applied  fields  in a bilayer graphene
\cite{cao18} and the one reported two years before in several bulk
samples with internal interfaces \cite{pre16}. Figure~\ref{tra}
shows the two sets of normalized resistance data for a better
comparison. As discussed in \cite{pre16} the background resistance
in Fig.~\ref{tra}(a)   is simply because the voltage electrodes do
not touch the interface(s) of interest. Both transitions show some
similarities worth mentioning,  in spite of the two orders of
magnitude difference in temperature.  Namely, there is not a
simple shift of the transition to lower temperatures with magnetic
field, but a small applied field already prevents  a complete
superconducting path between the voltage electrodes. Although  in
the case of the bulk graphite sample (Fig.~\ref{tra}(a)) one would
tend to explain this fact by an extra magnetoresistance coming
from the background resistance, this does not seem to be the
reason for the bilayer device.

Whereas in the case of the bulk graphite  sample the transition
hardly shifts to lower temperatures under an applied magnetic
field (in the measured field region), the broadness of the
transitions in the bilayer prevents a clear determination of a
temperature dependent upper critical field $B_{c2}(T)$. The
overall results suggest the existence of granular
superconductivity in both samples. In other words, neither in the
bilayer nor in the internal interface(s) of the bulk sample a
homogeneous superconducting region between the voltage electrodes
exists but superconducting patches. A magnetic field influences
the (Josephson) coupling between those patches and therefore no
zero resistance path between the electrodes remains. In spite of
granular superconductivity, at low enough fields permanent current
paths can exist, the
 reason for flux trapping \cite{sti18} and the remanent resistance
 observed  after removing the applied field \cite{pre16}. Future studies should clarify whether
the granular superconductivity is intrinsic  or extrinsic due to
defects (or inhomogeneous doping) at  the interfaces or in the
bilayer graphene.

The Josephson response between superconducting  patches that exist
within the embedded interfaces in bulk graphite samples, can be
measured  by depositing electrodes directly at the edges of the
interfaces, as has been done using TEM lamellae in \cite{bal13}.
Let us compare the results published in 2013  of one of those TEM
lamellae with one of the bilayer graphene devices published
recently. Figure~\ref{lam}(a) shows the temperature dependence of
the resistance of device M2, see Fig.1(b) and (e) in \cite{cao18},
and of the voltage (measured at constant current) of the lamella
L3, see Fig.3(c) and (f) in \cite{bal13}. Note first that the
transition observed in the lamella does not represent necessarily
the critical temperature of the superconducting patches but  the
 temperature where the Josephson coupling gets robust enough to influence the measured voltage.
 The Josephson response was identified  by measuring  current-voltage characteristics in
 both samples  shown in Figure~\ref{lam}(b) at two temperatures.
 The Josephson characteristics curves measured in  sample L3  at higher temperatures \cite{bal13} are also similar to those
 measured in the M2 device \cite{cao18} and can be very well understood following the Ambegaokar and Halperin model  \cite{amb69}.
Furthermore, transport measurements on different thin graphite
samples revealed that under a large enough applied electric field
near surface regions undergo a superconducting-like transition at
$T \sim 17~$K \cite{bal14}.  The  similarities between the
reported measurements in \cite{bal13,cao18} make any further
comment superfluous. Recently, using  point contact spectroscopy
at the surface of a graphite sample, local superconductivity was
found \cite{arn18}. The BCS-like features reveal a magnetic field
dependent energy gap with a critical temperature of 14~K
\cite{arn18}. Already in 1992 similar BCS-like features were found
by STM measurements at low temperatures, localized at certain
unknown regions at the surface of HOPG sample, without attracting
the attention of the scientific community  \cite{agr92}. Future
experiments should try to localize and characterize  the
interfaces in bulk graphite samples, which show  high critical
temperatures in order to hopefully start their difficult but
necessary production.

Finally,  apart from the interesting  cases of superconductivity
found in oxides films, semiconducting superlattices  and ultra
thin films (see, for example, \cite{moh14,bos16,wan16} and refs.
therein), there are some  examples in literature on  the existence
of superconductivity at certain interfaces that we would like to
note. Superconductivity has been found at the interfaces of pure
Bi (a material with some similarities to graphite) and BiSb
bicrystals up to critical temperatures $ \lesssim 21~$K
\cite{gip92,mun95,mun07,mun08}. Moreover, dislocations at certain
interfaces of semiconducting superlattices are thought to trigger
superconductivity up to 6~K \cite{fog01,fog06}, an idea that has
been also proposed for graphite~\cite{esqarx14}.  Also
strain-induced superconductivity at interfaces of semiconducting
layers has been treated theoretically based on the influence of
partial flat-bands \cite{tan014}.

Acknowledgements: We are grateful to A. Ballestar,  M. Stiller and
J. Barzola-Quiquia for discussions and for share their unpublished
magnetic force microscopy data on TiO$_2$ films are gratefully
acknowledge. We acknowledge the help of C. Precker for
digitalizing the data of Fig.1(b).


%

\end{document}